\documentclass[pra,twocolumn,showpacs,superscriptaddress,floatfix, reprint]{revtex4-1}

\usepackage[OT1]{fontenc}
\usepackage[usenames,dvipsnames]{color}
\usepackage[latin1]{inputenc}
\usepackage[english]{babel}
\usepackage{graphicx}
\usepackage{color}
\usepackage[Gray,squaren]{SIunits}
\usepackage{xspace}
\usepackage{ulem}
\usepackage{epstopdf}
\usepackage{amssymb,amsmath,verbatim,ulem}
\usepackage{dcolumn}
\usepackage{bm}
\usepackage{braket}
\usepackage{tabularx}
\usepackage{rotating}
\usepackage{hyperref}
\usepackage[caption=false]{subfig}
\usepackage{lipsum}

\newcommand{\RNum}[1]{\uppercase\expandafter{\romannumeral #1\relax}}

\begin{document}
\title{Four wave mixing based generation and control of light pulse} 
\author{Nawaz Sarif Mallick}
\email{nawaz@iitg.ac.in}

\author{Tarak Nath Dey}
\email{tarak.dey@iitg.ac.in}
\affiliation{Department of Physics, Indian Institute of Technology
Guwahati, Guwahati, Assam 781039, India}

\date{\today}

\begin{abstract}
We present an efficient scheme for the generation and control of a degenerate four-wave mixing (FWM)
signal in a $N$-type inhomogeneously broadened $^{85}$Rb atomic system. We observe the
propagation dynamics of the generated FWM signal along with the probe pulse under the condition of
Electromagnetically Induced Transparency. The FWM signal clones the temporal shape of the
probe pulse and travels through the medium without changing its shape and intensity.
We have also shown that a time dependent control field permits the storage and retrieval of these
optical signals without losing their identity. 
This work allows us to generate, control, store and retrieve FWM signal of any arbirary shape.
\end{abstract}

\maketitle
\section{Introduction} 
Atomic coherence induced by coherent light matter interaction has significant role for precise
control of the optical property of a medium. The observation of atomic coherence in an atomic
system leads us to uncover many spectacular optical effects. Among the most well known effect
is Electromagnetically Induced Transparency (EIT) \cite{RMP2005,GSA2012} in which an opaque
medium becomes eminently transparent for the probe field with the support of a control field.
Diverse applications such as slowing and stopping of light \cite{Juzeli2002,Liu2017,Arsenovi2018},
coherent storage and retrieval of light \cite{Patnaik2004,Xu2016,Rui2016,Katz2018},
Rydberg blockade induced interactions \cite{JDP2010},
diffraction control and guiding of light \cite{Evers2011, Cui2017}, structured beam generation \cite{Radwell,Sharma}, etc.
have been documented using EIT.
In multi-level atomic system, EIT enhances the nonlinear susceptibility
which conducts us to use nonlinear optical regime in the investigation many nonlinear optical phenomena such as,
Kerr nonlinearity \cite{Hamedi2015,Rajitha2015,Wu2015}, self-phase modulation (SPM) \cite{Wong2001,Kasparian2018},
cross-phase modulation (XPM)\cite{Sun2007,Dmochowski2016}, four-wave mixing (FWM) \cite{Lezama2000,Kang2004,Jeong2016}, etc.
In FWM process, three electro-magnetic fields interact in a nonlinear optical system and generate
electro-magnetic field with a new frequency.
Numerous experiments have been carried out to demostrate the enhanced FWM process
in multilevel atomic systems \cite{Epple2012,Lee2016,ZiYu2017,Juo2018,Zubairy2018}.
FWM process using EIT has been observed in both cold \cite{Lezama2000,Danielle2004,Kang2004,Lee2016,ZiYu2017,Juo2018,Zubairy2018} and
room temperature \cite{Li1996,Epple2012} atomic system.
Besides FWM process, enhanced higher order multi-wave mixing processes have been studied \cite{Zhang2007,Nie2008}. 
Hoonsoo $et$ $al.$ experimentally demonstrated EIT based six-wave mixing (SWM) signal in a $N$-type cold atomic system
\cite{Hoonsoo2004}. Recently an experimental observation of FWM signal in a $N$-type cold atomic system has been studied
by Chang-Kai $et$ $al.$ at low light levels \cite{Chang-Kai2014}.

The four-level system not only generates new signal but also it permits the generated signal to
propagate through the nonlinear medium along with the probe field under the condition of EIT \cite{Kang2004,Chang-Kai2014}.
The effect of enhanced nonlinearity on these propagating signals through the atomic medium remains unexplored.
Also the shape of the generated FWM signal and the progression during its propagation through the inhomogeneously broadened medium
is not completely elucidated.
In this paper, we theoretically investigate all those questions
in a simple four-level $N$-type atomic configuration as shown in Fig \ref{Fig1}.
A weak probe field and a strong control field resonantly drive the system.
In order to efficiently generate a FWM signal,
it is essential that the phase-matching condition, $2\vec{K}_p=\vec{K}_c+\vec{K}_g$ is strictly fulfilled \cite{Epple2012}.
Here we conceive this by using a collinear geometry, which generates the FWM signal 
in the same direction as the probe field propagates. The steady state of the optical Bloch equations 
are numerically solved to study the atomic coherence created by the nonlinear atom-field interaction.
The numerical result clearly shows gain in the FWM spectrum which indicates the possibility of
generation of a $\sigma^+$-polarised ($\Delta m=+1$) signal. The frequency of the signal,
$\omega_g$ related to the frequency of the probe field, $\omega_p$ and the frequency of the control field,
$\omega_c$  by $\omega_g=2\omega_p-\omega_c$.

Apart from the generation and control of the FWM signal, the storage and retrieval of this signal along with the
probe has captivated enormous attention due to its potential application as an optical memory
\cite{Camacho2009,Zaremba2002,Irina2011,Jinghui2013}.
Recently many experiments demonstrate  simultaneous storage and retrieval of both the signals in multi-level
atomic system \cite{Irina2011,Jinghui2013}. But the correlation between the input and the output pulse shape has not yet been explored in details.
It is noticeable from the previous demonstrations that the retrieved FWM signal does not preserve its predifined
shape and also its intensity is reduced considerably. We overcome these limitations by considering a much
simpler degenerate atomic system in which only one probe and one control field is interacting nonlinearly.
In this paper we demonstrate the shape preserving storage and retrieval of the FWM signal without
compromising its intensity. We also find that the storage and retrieval process is robust with
both the adiabatiac and non-adiabatic switching of the control field.

The arrangement of the article is as follows. In section \ref{Model}, we configure the physical model and describe
the system using a semiclassical theory.
In section \ref{Dynamical}, we discuss the dynamical equations of motion for the $N$-type system using Liouville's equation.
In section \ref{Pulse}, we derive the pulse propagation equations for the optical fields.
In the next section \ref{Generation}, we investigate the generation and control of the FWM signal.
In section \ref{Storage}, storage and retrieval of the FWM signal is demonstrated.
In section \ref{analysis}, we derive the analytical expression of the nonlinear coherence under weak probe approximation
in order to explain the FWM scheme.
Finally in section \ref{Conclusion}, we briefly conclude our work.
\section{THEORETICAL MODEL} 
\subsection{Model Configuration}
\label{Model}
\begin{figure}[t]
\includegraphics[scale=0.35]{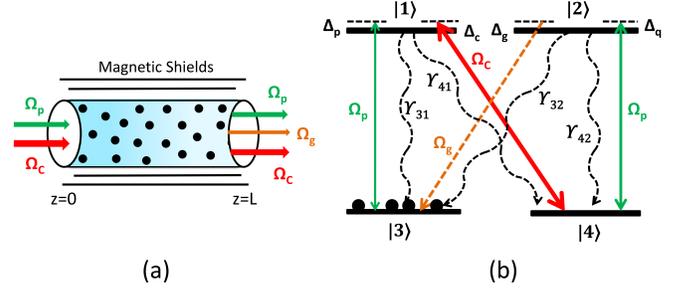}
\caption{(a) A simple block diagram of the the model system. (b) Schematic representation of the four-level $N$-type
atomic system. We use degenerate hyperfine sublevels of
$5^2 S_{\frac{1}{2}}$ and $5^2 P_{\frac{1}{2}}$. Two metastable ground states are defined as
$\ket 3=\ket {F=3,m_F=1}$ and $\ket 4=\ket {F=3,m_F=2}$.
Two excited states are defined as $\ket 1=\ket {F^{'}=2,m_{F^{'}}=1}$
and $\ket 2=\ket {F^{'}=2,m_{F^{'}}=2}$.}
\label{Fig1}
\end{figure}

In this work, the four wave mixing mechanism has been exploited for the generation and control of an optical signal.
The model system consists of an inhomogeneously broadened four level $^{85}$Rb atomic system interacting with two co-propagating optical fields as shown in Fig \ref{Fig1} . 
The atomic transitions $\ket1\leftrightarrow\ket3$ and $\ket2\leftrightarrow\ket4$ are coupled by a $\pi$-polarised ($\Delta m=0$) probe field of frequency $\omega_p$, whereas $\sigma^-$-polarised ($\Delta m=-1$) control field with frequency $\omega_c$ couple $\ket1\leftrightarrow\ket4$ transition. 
These optical fields are defined as
\begin{equation}\label{eq:field}
\vec{E}_{j}(z,t)=\hat e_{j}\mathcal E_{0j}(z,t)e^{i(k_j z-\omega_j t)}+c.c.,	
\end{equation}
where $\mathcal E_{0j}(z,t)$ is the space-time dependent amplitude, $k_{j}=\omega_{j}/c $ is the propagation constant along $z$-direction and $\hat e_{j}$ is the polarisation unit vector of the optical field. 
The subscript, $j\in \{p,c\}$ indicates the probe field and control field respectively.
The Hamiltonian describing the interaction between the four atomic level system with the two optical fields under electric-dipole approximation is as given below
\begin{equation}\label{eq:1}
\begin{aligned}
H^{'}=&\hbar \omega_{13}\ket1\bra1+\hbar(\omega_{13}-\omega_{14})\ket4\bra4\\
+&\hbar(\omega_{13}-\omega_{14}+\omega_{24})\ket2\bra2-\hbar\Omega_{p}e^{-i\omega_{p}t}\ket1\bra3\\
-&\hbar\Omega_{c}e^{-i\omega_{c}t}\ket1\bra4-\hbar\Omega_{p}e^{-i\omega_{p}t}\ket2\bra4 + h.c.,\\
\end{aligned}
\end{equation}
where the Rabi frequencies of the probe and control are defined by
\begin{equation}\label{eq:rabi}
\Omega_p=\frac{\vec{d}_{13}.\hat e_p}{\hbar} \mathcal E_{0p} \quad \textrm{and} \quad \Omega_c=\frac{\vec{d}_{14}.\hat e_c}{\hbar} \mathcal E_{0c}.
\end{equation}
The dipole moments for the atomic transitions between states $\ket1\leftrightarrow\ket3$, $\ket2\leftrightarrow\ket4$ and $\ket1\leftrightarrow\ket4$ are denoted by $\hat {d}_{13}$, $\hat {d}_{24}$, and $\hat {d}_{14}$, respectively.
Note that $\hat {d}_{13}$ = $\hat {d}_{24}$ according to the Clebsch-Gordan coefficients of the considered level scheme.
We perform the following unitary transformation in order to remove the explicit time dependence from the Hamiltonian
\begin{equation}
H=U^\dagger H^{'} U -i\hbar U^\dagger \frac{\partial U}{\partial t},
\end{equation}
where $U$ is defined as
\begin{equation}
U=e^{-i\left(\omega_{p}\ket1\bra1 + (\omega_{p}-\omega_{c})\ket4\bra4 + (2\omega_{p}-\omega_{c})\ket2\bra2\right)t}.
\end{equation}
Now the transformed Hamiltonian $H$  takes the following form
\begin{equation}\label{eq:2}
\begin{aligned}
H=-&\hbar\Delta_p\ket1\bra1-\hbar(\Delta_p-\Delta_c)\ket4\bra4\\
-&\hbar(\Delta_p-\Delta_c+\Delta_q)\ket2\bra2-\hbar\Omega_p\ket1\bra3\\
-&\hbar\Omega_c\ket1\bra4-\hbar\Omega_p\ket2\bra4 + h.c.,\\
\end{aligned}
\end{equation}
where $\Delta_p=\omega_p-\omega_{13}$, and $\Delta_q=\omega_p-\omega_{24}$  are the detunings of the probe field with $\ket1\leftrightarrow\ket3$ and $\ket2\leftrightarrow\ket4$ transitions, respectively.
The detuning due to the control field corresponding to  $\ket1\leftrightarrow\ket4$ transition is given by $\Delta_c=\omega_c-\omega_{14}$.
\subsection{Dynamical Equations}
\label{Dynamical}
We use Liouville's equation to account for the various radiative and non-radiative decay processes of the atomic system.
The decay of the atomic system is caused by various mechanisms such as flight-through broadening, population exchange, and atom-atom and atom-wall collisions.
To govern the response of the atomic populations and coherences of the four-level atomic system, the following density matrix equations are employed 

\begin{equation}
\dot \rho=-\frac{i}{\hbar}[H,\rho]+ \mathcal{L}_\rho 
\end{equation}

where the second term represents the decay processes that can be determined by
\begin{align}
\mathcal{L}\rho = \mathcal{L}_{r}\rho+\mathcal{L}_{c}\rho
 \label{decay1}
\end{align}
with
\begin{align}
\mathcal{L}_{r}\rho = &-\sum\limits_{i=1}^2\sum\limits_{j=3}^4 \frac{\gamma_{ji}}{2}\left(\ket i \bra i \rho-2\ket j \bra j \rho_{ii}+\rho\ket i \bra i \right)\,.
\label{decay2}
\end{align}
The spontaneous decay rates from the excited state $\ket i$, $(i \in 1,2)$ to the ground state $\ket j$, $(j \in 3,4)$ are denoted by $\gamma_{ji}$  in Eq.(\ref{decay2}).  
The dephasing in the ground state [$\mathcal{L}_{c}\rho$ in Eq.(\ref{decay1})] is due to collision at a rate $\gamma_c$.
Now the dynamics of the model system can be obtained in the following form

\begin{equation}\label{eq:r}
\begin{aligned}
\dot\rho_{11}&=-(\gamma_{31}+\gamma_{41})\rho_{11}+i\Omega_p\rho_{31}+i\Omega_c\rho_{41}-i\Omega_p^*\rho_{13}\\
&-i\Omega_c^*\rho_{14},\\
\dot\rho_{12}&=-[i(\Delta_q-\Delta_c)+\frac{1}{2}(\gamma_{31}+\gamma_{41}+\gamma_{32}+\gamma_{42})]\rho_{12}\\
&+i\Omega_p\rho_{32}+i\Omega_c\rho_{42}-i\Omega_p^*\rho_{14},\\
\dot\rho_{13}&=[i\Delta_p-\frac{1}{2}(\gamma_{31}+\gamma_{41})]\rho_{13}+i\Omega_p(\rho_{33}-\rho_{11})\\
&+i\Omega_c\rho_{43},\\
\dot\rho_{14}&=[i\Delta_c-\frac{1}{2}(\gamma_{31}+\gamma_{41})]\rho_{14}+i\Omega_p(\rho_{34}-\rho_{12})\\
&+i\Omega_c(\rho_{44}-\rho_{11}),\\
\dot\rho_{22}&=-(\gamma_{32}+\gamma_{42})\rho_{22}+i\Omega_p\rho_{42}-i\Omega_p^*\rho_{24},\\
\dot\rho_{23}&=[i(\Delta_p-\Delta_c+\Delta_q)-\frac{1}{2}(\gamma_{32}+\gamma_{42})]\rho_{23}\\
&+i\Omega_p(\rho_{43}-\rho_{21}),\\
\dot\rho_{24}&=[i\Delta_q-\frac{1}{2}(\gamma_{32}+\gamma_{42})]\rho_{24}+i\Omega_p(\rho_{44}-\rho_{22})\\
&-i\Omega_c\rho_{21},\\
\dot\rho_{33}&=\gamma_{31}\rho_{11}+\gamma_{32}\rho_{22}+i\Omega_p^*\rho_{13}-i\Omega_p\rho_{31}\\
\dot\rho_{34}&=-[i(\Delta_p-\Delta_c)+\gamma_{c}]\rho_{34}+i\Omega_p^*\rho_{14}-i\Omega_p\rho_{32}\\
&-i\Omega_c\rho_{31},\\
\dot\rho_{44}&=-(\dot\rho_{11}+\dot\rho_{22}+\dot\rho_{33})\\
\dot\rho_{ij}&=\dot\rho_{ji}^*
\end{aligned}
\end{equation}
where the overdot stands for the time derivative and star $(*)$ denotes the complex conjugate.
For an inhomogeneously broadened medium, frequency shift due to finite atomic motion is inhibited.
In presence of Doppler shift, the probe and control field detuning are modified to
$\Delta^{'}_{p}=\Delta_{p}-k_{p}v$, $\Delta^{'}_{q}=\Delta_{p}-k_{p}v$, $\Delta^{'}_{c}=\Delta_{c}-k_{c}v$, respectively.
The sign of frequency shift $\pm k_{i}v, (i\in{p,q,c})$ indicates the motion of atoms either counter propagating or co-propagating.
Therefore, these effects can be incorporated into the equations of motion (\ref{eq:r}) by taken into account the avearging over a Maxwell velocity distribution.
Hence the velocity averaging of the atomic coherences $\langle\rho_{ij}(z,t)\rangle$ can be expressed as
\begin{equation}
\langle\rho_{ij}(z,t)\rangle= \int \rho_{ij}(z,v,t)\mathcal{P}(kv)d(kv),
\end{equation}
where $\mathcal{P}(kv)d(kv)$ is probability that an atom has a velocity between $v$ and $v + dv$ and obeyed by the Maxwell-Boltzmann velocity distribution
\begin{equation}\label{eq:distribution}
\mathcal{P}(kv)d(kv)=\frac{1}{\sqrt{2\pi D^{2}}}  e^{-\frac{(kv)^2}{2D^2}} d(kv).
\end{equation}
The Doppler line width D  is given by $D=\sqrt{k_{B}T\nu^2_{c}/Mc^2}$, where $M$ is the atomic mass, $k_{B}$ is the Boltzmann constant and $T$ is the thermal equilibrium temperature. 
At room temperature (T=300K), the Doppler width, $D$ is 37$\gamma$ for $^{85}$Rb atoms.
\subsection{Pulse Propagation Equations}
\label{Pulse} 
In this section, we use Maxwell's equations to govern the spatio-temporal evaluation of an optical field through a nonlinear medium.
The wave equation for the probe, control and generated fields can be expressed as
\begin{equation}\label{eq:wave_eq}
\left(\nabla^2+\frac{1}{c^2}\frac{\partial^2}{\partial t^2}\right)\vec{E}=\frac{4\pi}{c^2}\frac{\partial^2 \vec{\mathcal P}}{\partial t^2},
\end{equation}
where $\vec{E}=\vec{E}_p+\vec{E}_c+\vec{E}_g$ is the total field and its induced polarization  is $\vec{\mathcal P}$.
The source term  $\vec{\mathcal P}$ that appears in the right hand side of Eq.(\ref{eq:wave_eq}) is the origin of  linear and non-linear response of the medium.
The macroscopic polarization $\vec{\mathcal P}$ can be defined in terms of the atomic coherences as
\begin{equation}\label{eq:polarization}
\begin{aligned}
\vec{\mathcal P}&=\mathcal N((\vec{d}_{13}\rho_{13}+\vec{d}_{24}\rho_{24})e^{-i\omega_p t}\nonumber\\
&+\vec{d}_{14}\rho_{14}e^{-i\omega_c t}+\vec{d}_{23}\rho_{23}e^{-i\omega_g t}+c.c.,)
\end{aligned}
\end{equation}
where $\mathcal{N}$ is the atomic density. 
Note that the atomic density under consideration is weak, otherwise local field correction need to be taken into account.
Under slowly-varying envelope approximation (SVEA), Eqs.(\ref{eq:wave_eq}) can be reformulated as
\begin{equation}\label{eq:14}
\begin{aligned}
&\left(\frac{\partial }{\partial z}+\frac{1}{c}\frac{\partial }{\partial t}\right)\Omega_p=i\eta_p\left\langle \rho_{13}(z,t)+\rho_{24}(z,t) \right\rangle,\\
&\left(\frac{\partial }{\partial z}+\frac{1}{c}\frac{\partial }{\partial t}\right)\Omega_c=i\eta_c\langle\rho_{14}(z,t)\rangle,\\
&\left(\frac{\partial }{\partial z}+\frac{1}{c}\frac{\partial }{\partial t}\right)\Omega_g=i\eta_g\langle\rho_{23}(z,t)\rangle.\\
\end{aligned}
\end{equation}
The coupling constant for the probe, control and generated fields are defined as
$\eta_p=\eta_c=\eta_g=\eta=3 \mathcal N \gamma \lambda^2/8 \pi$. 
For simplicity, we consider $\gamma=\gamma_{j1}=\gamma_{j2}$ $(j \in 3,4)$, and $\lambda_{13}=\lambda_{14}=\lambda_{23}=\lambda$.
The angular bracket denotes a statistical average over the velocity distribution of the atom.
We introduce the following co-moving coordinate system in order to perform numerical computation:
\begin{equation}
\tau=t-\frac{z}{c},\hspace{0.25cm}\xi=z.
\end{equation}
 Therefore, the expressions within the round bracket of Eq.(\ref{eq:14}) can be easily substituted by $\partial/\partial \xi$ in  the frame of moving coordinate system,
Subsequently, simultaneous solutions of Bloch Eqs.(\ref{eq:r}) and Maxwell's Eqs.(\ref{eq:14})  in space-time coordinate explore the dynamical 
progression of the optical fields inside the medium.

\section{Generation and control of FWM signal}
\label{Generation} 
In this section, we proceed with numerical simulation of Maxwell-Bloch equations to reveal
the generation of optical field  and its control by exploiting FWM mechanism.
We have adopted Cash-Karp Runge-Kutta method in order to solve coupled partial differential equations.
We begin with lay out of the initial and boundary conditions of the atoms and fields.
The time dependent envelope of the probe field at the entry face of the medium can be written as
\begin{equation}\label{eq:probeshape}
\Omega_p(\xi=0,\tau)=\Omega_p^0 e^{-(\frac{\tau-\tau_0}{\sigma_p})^2},
\end{equation}
where $\Omega_p^0$, $\sigma_p$ and $\tau_0$ are the amplitude, temporal width and peak location of the Gaussian shaped probe field, respectively.
In our entire computation, we have assumed that the amplitude of control field ($\Omega^{0}_{c}=3.0\gamma$) is larger than the amplitude of the probe field ($\Omega^{0}_{p}=0.1\gamma$) and 
subsequently the dynamical evolution of control field can be neglected.
Initially all atoms are occupied in the $\ket{3}$ state whereas all other states are unoccupied for $\xi\in(0,L)$.
\begin{figure}
\includegraphics[scale=0.725]{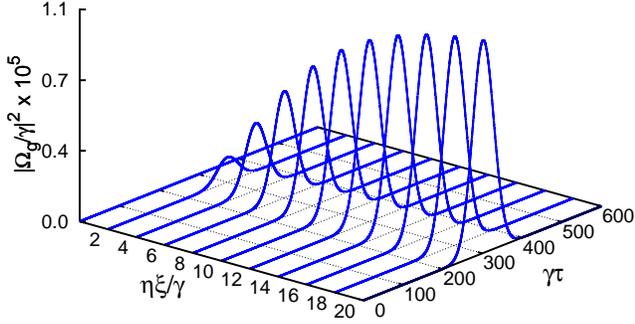}
\caption{Propagation dynamics of the generated FWM signal as function of position and time.
The parameters are $\Omega_p^0=0.1\gamma$, $\Omega_c^0=3.0\gamma$, $\mathcal N=1\times 10^{12}$atoms/cm$^3$, $\lambda=7.95\times 10^{-5}$cm,
$\Delta_p=\Delta_c=\Delta_{q}=\Delta_g=0$,
$\sigma_p=75/\gamma$, $\gamma\tau_0$=300, $\gamma_{c}=0$.}
\label{Fig2}
\end{figure}

\begin{figure}
\includegraphics[scale=0.35]{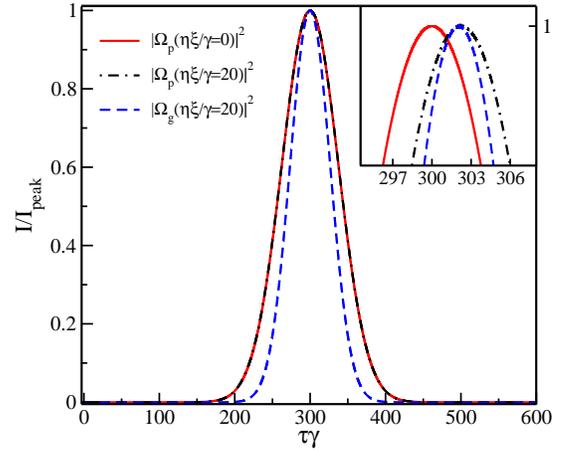}
\caption{Normalised intensity profile, $I/I_{peak}$ of the Gaussian shaped probe pulse and FWM signal at
input ($\eta\xi/\gamma=0$) and output ($\eta\xi/\gamma=20$) boundary of the medium. Temporal peak position of the
output pulses are shifted at $\gamma\tau$=300. Inset figure shows the actual peak location.   
All other parameters are same as in Fig. (\ref{Fig2}).}
\label{Fig3}
\end{figure}

Our first finding on the generation of optical pulse in the presence of a continuous wave (cw) control field is shown in Fig.\ref{Fig2} .
Figure (\ref{Fig2}) shows the temporal variation of generated field at different propagation distances.
It is clear from  Fig. (\ref{Fig2}) that the amplitude of the generated field increases gradually and takes the probe field shape while propagating along the medium.
Note that temporal shape of the generated field remains unchanged after it attains saturation intensity.
For the sake of generality, we compare the temporal shape and width of the generated field with the probe field as in Fig. \ref{Fig3}.
It can be seen that the time dependent envelope of the generated field is same as the input probe field, except its width is narrowed by a factor of $\sqrt{2}$.
We also notice that the temporal shapes of both generated and probe fields propagate through the medium without absorption and distortion.
Therefore, only the cw control field efficiently clones the temporal shape of probe field to generate signals and protect them during propagation.

\begin{figure}
\includegraphics[scale=0.35]{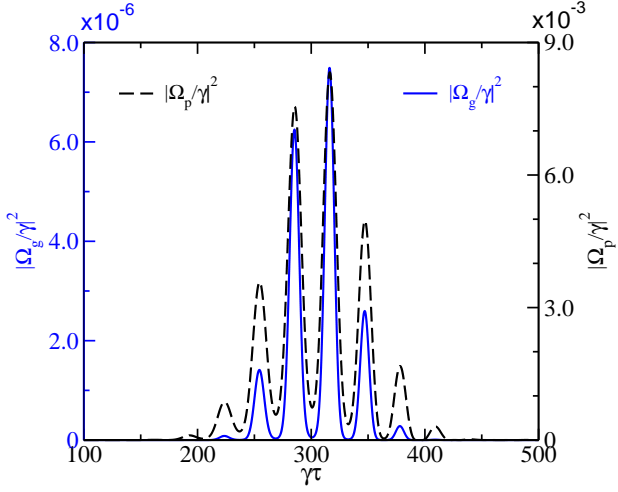}
\caption{Intensity profile of the amplitude modulated Gaussian shaped probe pulse and generated FWM signal
at output ($\eta\xi/\gamma=20$) boundary of the medium.
The parameters for the probe pulse are $\Omega_p^0=0.06\gamma$, $m_a=0.65$, $w_m=\gamma/5$, $\sigma_p=75/\gamma$ and $\gamma\tau_0$=300.
Other parameters are same as in Fig. (\ref{Fig2}).}
\label{Fig4}
\end{figure}

In order to prove the efficiency and the robustness of the model system, we next consider input probe field to be an amplitude modulated Gaussian pulse.
For this purpose, the input envelope for the probe field is expressed as follows:
\begin{equation}\label{eq:dprobeshape}
\Omega_p(\xi=0,\tau)=\Omega_p^0\left(1+m_a\cos \omega_m t \right) e^{-(\frac{\tau-\tau_0}{\sigma_p})^2}
\end{equation}
where $m_a$ and $\omega_m$ are termed as the depth of modulation and frequency of the modulating signal, respectively. 
In practice,  $\omega_m$ is small compared to the carrier frequency $\omega_p$ of the probe pulse.
Figure \ref{Fig4} depicts the intensity profile of the probe and generated fields as a function of $\gamma\tau$ at propagation distance $\eta\xi/\gamma = 20$.
This figure confirms precisely that the generated FWM signal gets its shape from the envelope of the probe field and propagates as a shape preserving pulse.
However the time resolution of the generated envelope is higher as compared to the probe pulse.

\section{Storage and Retrieval of EM radiations}
\label{Storage}
\begin{figure}
\includegraphics[scale=0.725]{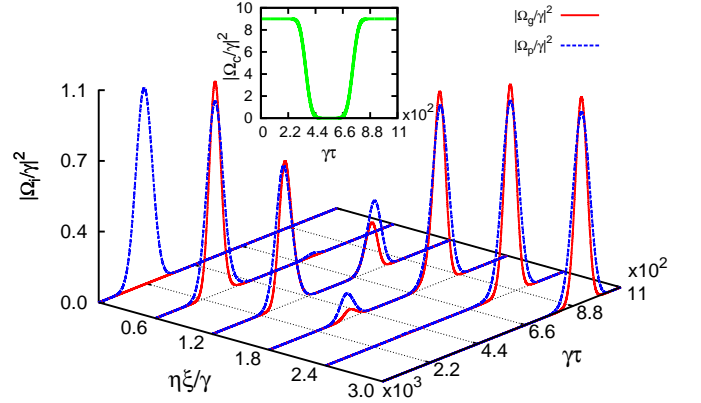}
\caption{Storage and retrieval of Gaussian probe pulse (dashed blue line, $i=p$) and generated FWM signal (solid red line, $i=g$)
are demonstrated. We multiply $10^{2}$ and $10^{5}$ with the probe field intensity ($|\Omega_p/\gamma|^2$) and FWM signal intensity
($|\Omega_g/\gamma|^2$) respectively. Inset figure shows intensity profile the control field (solid green).
The parameters are $\sigma_c=180/\gamma$, $\gamma\tau_c=550$, $\alpha=4$ and all other parameters are same as in Fig. (\ref{Fig2}).}
\label{Fig5}
\end{figure}

In the previous section, we have demonstrated how the time independent control field efficiently generates and controls the propagaton of FWM signal.
Here we explore the dynamics of  the generated field in presence of time dependent profile of control field.
We address this issue by considering the temporal profile of control field to be of inverted super-Gaussian shape which is defined as:
\begin{equation}
\Omega_c(\xi=0,\tau)=\Omega_c^0 \Big[1-e^{-\big(\frac{\tau-\tau_c}{\sigma_c}\big)^{\alpha}}\Big],
\end{equation}
where the parameter $\alpha$ regulates the rapidity of the switching action of the control field. 
Switching off and on of control field intensity in time domain holds the key of storage and retrieval process.
Figure \ref{Fig5} displays the spatio-temporal characteristics of probe and FWM fields. 
The initial profile of the probe field  is chosen as a Gaussian. 
As seen from Fig. \ref{Fig5} the probe field intensity gradually diminishes due to dynamical reduction of control intensity.
Simultaneously FWM signal is formed and its spatio-temopral evolution follows the same dynamical behaviour as the probe field.

The control field not only generates the FWM signal but also enables the storage and retrieval of the signal along with the
probe pulse by temporal variation of its own intensity.
The intensity lowering (rising) to zero (maximum) with time produces switching off (on) of the control field as shown in inset of Fig.\ref{Fig5}.
The spatiotemporal evolution of probe and generated signals in the presence of super Gaussian shaped control field is shown in Fig \ref{Fig5}.
The signal and probe fields are depicted by solid red and dashed blue lines, respectively.
Figure \ref{Fig5} shows that the generation of signal field is accomplished within a  short length of medium in the presence of both control and probe fields.
It is also evident from Fig.\ref{Fig5} that storage of signal and probe fields start as soon as both peaks experience the falling intensity of the control field.
Hence the intensities of both signal and probe fields are gradually stored by adiabatic switching off the control field during the propagation through the atomic medium.
The storing of the fields can be completed by making intensity of control field zero. 
The stored pulses can be retrieved on demand by switching on the control field.
The retrieval process of stored fields can be initiated by switching on of the control field at a later time.
The maximum intensity of the control field leads to retrieval of both probe and signal from the medium without loss of generality.

\begin{figure}
\includegraphics[scale=0.35]{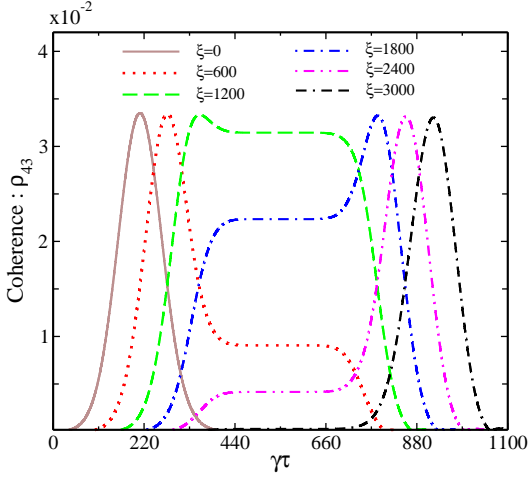}
\caption{The temporal profile of ground state atomic coherence, $\rho_{43}$ is plotted as
a function of time for different propagation distances. All parameters are same as in figure \ref{Fig5}.}
\label{Fig6}
\end{figure}

Next we investigate the ground state atomic coherence which is solely responsible for the storage and retrieval of electromagnetic radiation in an atomic medium \cite{LukinPRL2000}.
Subsequently we have plotted the atomic coherence $\rho_{43}$ as a function of time at different propagation distances
as shown in Figure \ref{Fig6}.
As depicted from Figure \ref{Fig6}, at entry face of the medium $\eta\xi/\gamma=0$, 
$\rho_{43}$ takes the temporal shape of probe field in presence of constant control. 
The atomic coherence attains its maximum value when control field is switched off. 
The probe and signal pulses are stored inside the medium in the form of ground atomic coherence
by switching off the control field.
This coherence ($\rho_{43}$) is well preserved inside the medium and  efficiently can be retrieved before it decays at a rate of $\Gamma_{43}$.
The atomic coherence starts generating the replica of the stored pulses after the control field is switched on.
The Raman scattering between stored atomic coherence and control field intensity produces back the stored signal.
Therefore, the atomic coherence plays the main role
behind the storage and retrieval of the probe as well as the FWM signal.

\section{ANALYSIS AND DISCUSSIONS}
\label{analysis}
\subsection{Perturbative Analysis} 
In this section, we derive an analytical expression for the atomic coherence which can successfully explain the generation of signal due to four wave mixing in four level atomic system.
The perturbative expression for the coherence and population is determined under a  weak probe approximation ($\Omega_p\ll\Omega_c$) that are correct in all orders for the control field of Rabi frequencies $\Omega_c$ and $2^{nd}$ order in probe Rabi frequencies $\Omega_p$.
The solutions of the density matrix equations can be approximated as 
\begin{equation}
\begin{aligned}
\rho_{ij}&=\rho_{ij}^{(0)}+\Omega_p\rho_{ij}^{(1)}+\Omega_p^* \rho_{ij}^{(2)}+\Omega_p^2 \rho_{ij}^{(3)}+|\Omega_p|^2 \rho_{ij}^{(4)}\nonumber\\
&+\Omega_p^{*2} \rho_{ij}^{(5)}\\
\end{aligned}
\end{equation}
where $\rho_{ij}^{(0)}$ describes the solution in the absence of $\Omega_p$ and $\rho_{ij}^{(k)}$, $k\in \{1,2,3,4,5\}$ is
the higher order solution in the presence of weak $\Omega_p$.
The steady-state value of the atomic coherence, $\rho_{23}$ can be expressed
by the following expression
\begin{equation}\label{eq:10}
\rho_{23}=\frac{\frac{i\Omega_p^2\Omega_{c}^*}{[i\Delta_{p^{'}}-\frac{\gamma_{31}+\gamma_{41}}{2}][i(\Delta_{p^{'}}-\Delta_{c^{'}}+\Delta_{q^{'}})-\frac{\gamma_{32}+\gamma_{42}}{2}]}}{\bigg[i(\Delta_{p^{'}}-\Delta_{c^{'}})-\Gamma_{34}+\frac{\Omega_{c}^2}{i\Delta_{p^{'}}-\frac{\gamma_{31}+\gamma_{41}}{2}}\bigg]}.
\end{equation}
The above expression corresponds to the four wave mixing in a system of four level atoms and produces a frequency  $\omega_g=2\omega_p-\omega_c$.
The induced nonlinear atomic polarization $\vec{P}^{NL}$ is expressed as
$\vec{P}^{NL} =\chi^{(3)} (\omega_g) \vec{E}_p^2 \vec{E}^*_c$ where $\chi^{(3)}(\omega_g)$ is a $3^{\textrm{rd}}$ order nonlinearity.
Under the Doppler-broaden, the nonlinear susceptibility  $\chi^{(3)}(\omega_g)$  can be written as
\begin{equation}
\langle \chi^{(3)} (\omega_g) \rangle=\frac{\mathcal N |\vec{d}_{23}| |\vec{d}_{13}|^2 |\vec{d}_{14}|}{\hbar^3 |\Omega_p|^2 \Omega^*_c} \langle \rho_{23} \rangle,
\end{equation}
where $\mathcal N$ is the atomic density.

\subsection{Nonlinear Susceptibility} 
\begin{figure}
\includegraphics[scale=0.35]{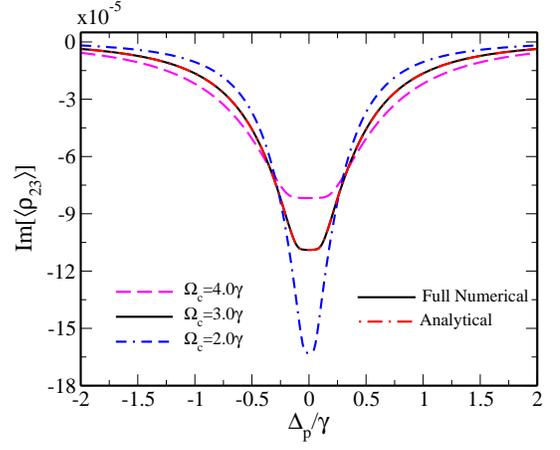}
\caption{Imaginary part of nonlinear Doppler averaged coherence, $\langle\rho_{23}\rangle$ as a function of $\Delta_p$ in units of $\gamma$
for two different approach and for various values of $\Omega_c$. The parameters are $\Omega_p=0.1\gamma$,
$\gamma=2\pi\times5.75\times10^6$ Hz, $D=37\gamma$, T=300K.}
\label{Figure7}
\end{figure}

We next study the medium with Kerr nonlinearity by using two different approaches: 
firstly we use weak-approximated analytical solution (\ref{eq:10}) to obtain response of the generated signal coherence. 
Secondly, we solve full  density matrix equations (\ref{eq:r}) numerically at steady state limit
to find the nonlinear susceptibility of signal.
The results from both approaches are in good agreement as shown in Fig.\ref{Figure7}. 
It ensures that a new optical field generation is possible even in the weak probe regime.
The gain of the generated signal can be enhanced by reducing the control field intensity.
This is possible due to the dark state ($\Omega_p \ll \Omega_c$) becoming feeble as the control field intensity is reduced significantly, as
causing the population leaves from dark state and populate in the excited states. 
Hence the efficiency of nonlinear signal generation can be enhanced with a suitable value of control field intensity.

\section{Conclusion} 
\label{Conclusion}
In conclusion, we have demonstrated an efficient generation and control of a degenerate FWM signal by optical cloning
in a $N$-type inhomogeneously broadened atomic system.
A strong control field and a weak probe field tailored the medium susceptibility with a gain profile which enable us to
generate the FWM signal. The control field can be utilized as a parameter to manage the efficiency of
the FWM signal. The propagation dynamics of the generated FWM signal and the effect of enhanced
non-linearity are studied in detail. The FWM signal clones the temporal shape of the probe field and travels through
the atomic medium along with the probe field without changing its shape and intensity.
This scheme allows us to generate an efficient FWM signal of any arbitrary shape. We have also demonstrated the coherent
storage and retrieval of the signal with the assistance of a time dependent profile of the control field.
The generation, control, storage and retrieval of such higher order nonlinear optical signals in multilevel atomic systems
can have potential applications in nonlinear optical spectroscopy, quantum information processing and optical communication systems.
\section*{Acknowledgments}
N. S. Mallick would like to thank MHRD, Government of India for financial support.



\bibliography{reference}
\end{document}